**UnespDataLens-RM: A Reference Model for Analytical Data Engineering with Governance, Quality, Provenance, and Reproducibility**

Ronaldo Celso Messias Correia[1], Douglas Francisquini Toledo[1], Camila Tolin Santos da Silva[1]
[1] São Paulo State University (UNESP), School of Technology and Sciences, Presidente Prudente, SP, Brazil

## Abstract

The growing reliance on data in analytical processes and evidence-based decision-making has reinforced the importance of Data Engineering in building pipelines capable of integrating, transforming, validating, and delivering data from heterogeneous sources. However, the reliability of analytical assets depends not only on data processing capabilities but also on mechanisms for governance, quality assurance, provenance, traceability, versioning, and reproducibility throughout their lifecycle. These responsibilities are commonly addressed by different models, frameworks, and operational practices, resulting in methodological fragmentation across the analytical data lifecycle. To address this gap, this article proposes UnespDataLens-RM, a technology-independent reference model that integrates technical-operational processes and cross-cutting capabilities within a unified structure for Analytical Data Engineering. The model aims to support the specification, organization, and evolution of analytical pipelines by incorporating governance, quality, provenance, traceability, and reproducibility from the design stage. Developed following the Design Science Research approach, UnespDataLens-RM comprises eight technical-operational modules, eight cross-cutting modules, complementary dimensions, and a formalized set of artifacts, metrics, and validation criteria. The resulting specification offers a conceptual and methodological framework for future instantiations and empirical evaluations of analytical pipelines designed to be more governable, documented, traceable, auditable, and reproducible.

## 1. Introduction

The ongoing digital transformation has significantly increased the production and availability of data across different sectors of society. Public and private institutions, research centers, and educational organizations continuously generate data from transactional systems, enterprise applications, digital platforms, sensors, connected devices, and cloud services. Data have consequently become valuable assets for analytical processes, knowledge generation, and evidence-based decision-making. However, data availability alone does not ensure the production of reliable information. Data quality is a fundamental requirement for effective use, encompassing dimensions that go beyond mere accuracy and directly influence fitness for purpose across different contexts (Wang; Strong, 1996). Poor data quality can also have significant organizational impacts and compromise processes that depend on those data (Redman, 1998). In contemporary environments, these challenges are further intensified by the heterogeneity of data sources and the complexity of the architectures and systems responsible for data storage, integration, and processing (Kleppmann, 2017; Reis; Housley, 2022).

In this context, Data Engineering plays a fundamental role in the design and maintenance of infrastructures capable of acquiring, integrating, transforming, storing, and delivering data for different analytical purposes. The evolution of these infrastructures has led to the emergence of architectures such as data lakes, lakehouses, and data mesh, while practices associated with DataOps and MLOps have increasingly emphasized automation, operationalization, monitoring, and the continuous evolution of data pipelines and analytical models (Armbrust et al., 2021; Dehghani, 2022; Kreuzberger; Kühl; Hirschl, 2023). These advances have substantially expanded the possibilities for building scalable and distributed data ecosystems, but they have also increased their technical and organizational complexity. Consequently, managing the lifecycle of the analytical assets produced by these pipelines has become increasingly important.

The literature has addressed these responsibilities from different perspectives. Studies on data quality, as discussed by Wang et al. (2024), highlight its multidimensional nature and the influence of technical, organizational, and contextual factors on data fitness for use. In the field of data governance, the definition of policies, responsibilities, decision-making structures, and control mechanisms has gained increasing importance as a means of managing data assets (Bliznák; Munk; Pilková, 2024). Data provenance and lineage, in turn, provide mechanisms for documenting the origin, dependencies, and transformations of assets throughout their lifecycle, thereby contributing to traceability and auditability. Computational reproducibility further requires the preservation of data, code, configurations, computing environments, and other information necessary to reconstruct analytical processes and results (Antunes; Hill, 2024). In addition, recent research on MLOps has consolidated practices related to automation, integration, deployment, monitoring, and continuous maintenance of machine learning systems, while also identifying challenges associated with standardization, scalability, and maturity (Zarour; Alzabut; Al-Sarayreh, 2025; Eken et al., 2025).

Although these contributions provide important foundations, they differ in purpose, scope, and level of abstraction. Similarly, process models such as KDD (Knowledge Discovery in Databases) and CRISP-DM (Cross-Industry Standard Process for Data Mining) structure activities related to knowledge discovery and data mining (Fayyad; Piatetsky-Shapiro; Smyth, 1996; Chapman et al., 2000), whereas CRISP-ML(Q) extends this perspective to the development of machine learning applications with explicit quality requirements (Studer et al., 2021). DAMA-DMBOK, by contrast, focuses primarily on data management and governance; data mesh emphasizes sociotechnical decentralization and a data-product orientation; and lakehouse proposes an architectural approach that combines capabilities traditionally associated with data lakes and data warehouses (DAMA International, 2017; Armbrust et al., 2021; Dehghani, 2022). These approaches are therefore complementary, but they do not necessarily provide a common structure for jointly organizing the responsibilities involved in Analytical Data Engineering.

The analysis of these approaches reveals a methodological fragmentation between the processes that produce analytical assets and the capabilities required to govern, assess, document, trace, and reproduce them. Within the set of approaches reviewed in this study, we did not identify a reference structure that explicitly relates technical-operational processes of Analytical Data Engineering to cross-cutting capabilities for governance, quality, metadata, provenance, versioning, reproducibility, and monitoring, while also associating these capabilities with design principles, artifacts, and evaluation mechanisms.

To address this problem, this article proposes UnespDataLens-RM, a reference model for Analytical Data Engineering designed to organize the lifecycle of analytical assets through the integration of technical-operational processes and cross-cutting capabilities. The model aims to provide a technology-independent structure capable of representing not only the operations responsible for data acquisition, integration, preparation, validation, storage, and delivery, but also the responsibilities required for governance, documentation, traceability, versioning, monitoring, and reproducibility.

The research is guided by the following question: How can a reference model be specified to integrate Analytical Data Engineering processes with governance, quality, provenance, and reproducibility capabilities within a common conceptual structure that supports the development of reliable, auditable, and reusable analytical pipelines? To address this question, UnespDataLens-RM was developed following the Design Science Research (DSR) approach, which is suitable for the construction and evaluation of artifacts intended to address relevant problems and generate prescriptive knowledge (Hevner et al., 2004; Peffers et al., 2007; Gregor; Hevner, 2013).

The resulting model is organized into eight technical-operational modules, which represent the main flow for producing analytical assets, and eight cross-cutting modules, which incorporate capabilities related to metadata, governance, provenance, versioning and reproducibility, analytical representations, feature engineering, monitoring, and intelligent support. The architecture is further complemented by dimensions designed to incorporate contemporary Data Engineering requirements and by a formalized set of artifacts, metrics, and validation criteria. This organization establishes explicit relationships between the processes that transform data and the mechanisms responsible for documenting, controlling, and monitoring their evolution.

The main contributions of this study are: (i) the proposal of a reference model for organizing the lifecycle of Analytical Data Engineering; (ii) the integration, within a common structure, of technical-operational processes and cross-cutting capabilities related to governance, quality, provenance, traceability, versioning, and reproducibility; (iii) the formalization of the model through design principles, modules, relationships, artifacts, metrics, and validation criteria; and (iv) the establishment of a technology-independent conceptual foundation capable of guiding future instantiations, domain specializations, and empirical evaluations.

The remainder of this article is organized as follows. Section 2 presents the conceptual background and related work, positioning the proposed model in relation to the main existing approaches. Section 3 describes the research method and the artifact development process according to DSR. Section 4 presents UnespDataLens-RM, including its design principles, formalization, architecture, and component organization. Section 5 discusses its contributions, relationships with existing approaches, potential applications, and limitations. Finally, Section 6 presents the conclusions and directions for future work.

## 2. Background and Related Work

Data Engineering provides the technical foundation for building and operating systems capable of acquiring, integrating, transforming, storing, and delivering data for different analytical purposes. Its scope encompasses decisions related to architecture, modeling, processing, orchestration, reliability, scalability, and maintenance of the infrastructures that support the data lifecycle (Kleppmann, 2017; Reis; Housley, 2022). While Data Science focuses predominantly on data exploration, modeling, and interpretation, Data Engineering provides the processes and infrastructure required for these activities to be performed consistently and sustainably.

In this article, the term Analytical Data Engineering is used to delimit the set of Data Engineering processes directly related to the production and maintenance of assets intended for analytical consumption. This perspective encompasses the lifecycle that begins with the identification and acquisition of data sources, proceeds through integration, transformation, validation, and storage, and culminates in the delivery of datasets, data products, and other structures intended for analysis, visualization, or computational modeling.

The evolution of data architectures has significantly changed how these processes are implemented. Architectures based on data warehouses have come to coexist with data lakes and, subsequently, with lakehouses, which seek to combine properties of data warehouses with the flexibility and scalability of data lakes (Armbrust et al., 2021). In parallel, data mesh introduced a decentralized sociotechnical perspective based on domain-oriented data ownership, data as a product, self-service infrastructure, and federated computational governance (Dehghani, 2022). These approaches provide alternative ways of organizing and implementing data ecosystems, but they do not, by themselves, constitute process models for the entire lifecycle of Analytical Data Engineering.

The systematization of analytical processes has important precedents in models developed for knowledge discovery and data mining. KDD structured knowledge discovery as a process comprising data selection, preprocessing, transformation, mining, and interpretation of results (Fayyad; Piatetsky-Shapiro; Smyth, 1996). Subsequently, CRISP-DM established an iterative process organized around business understanding, data understanding, data preparation, modeling, evaluation, and deployment (Chapman et al., 2000). With the expansion of machine learning into production environments, CRISP-ML(Q) incorporated quality assurance requirements into the development lifecycle of machine learning applications (Studer et al., 2021).

These models are well established, but they pursue objectives that differ from those of data governance frameworks and data architectures. CRISP-DM and CRISP-ML(Q) primarily structure activities associated with the development of analyses or models; DAMA-DMBOK organizes knowledge areas for data management; and lakehouse characterizes an architectural approach. Consequently, these proposals should not be interpreted as directly equivalent alternatives, but rather as complementary approaches.

This distinction also defines the concept of a reference model adopted in this study. A reference model is understood as an abstract, implementation-independent representation that organizes the elements, responsibilities, and relationships of a given domain, providing a common structure from which different instantiations can be developed (OASIS, 2006). From this perspective, UnespDataLens-RM is not intended to replace architectures, process models, or specialized frameworks. Instead, it establishes a structure within which such approaches can contribute to the implementation of specific responsibilities.

The literature analyzed in this study shows that the capabilities required to build reliable analytical ecosystems are distributed across approaches developed for different purposes. Process models emphasize activities related to knowledge discovery and the development of analytical solutions; management frameworks focus on governance, quality, and metadata; architectures define ways of organizing and storing data; and operational practices extend automation, versioning, and monitoring capabilities. The research gap addressed in this study therefore does not arise from the individual absence of these capabilities, but from their fragmentation across different conceptual, methodological, and operational levels.

To summarize this positioning, Table 1 compares representative approaches according to capabilities related to the scope of UnespDataLens-RM. Because these approaches differ in purpose and level of abstraction, the comparison is not intended to establish superiority among them. The classification only considers whether a given capability is explicitly addressed in the specification of each approach: Yes indicates explicit coverage; Partial indicates limited, indirect, or scope-restricted treatment; and No indicates that the capability is not explicitly specified as part of the approach.

**Table 1.** Coverage of capabilities across approaches related to UnespDataLens-RM

| Capability | CRISP-DM | CRISP-ML(Q) | DAMA-DMBOK | UnespDataLens-RM |
|---|---|---|---|---|
| Analytical production process | Yes | Yes | No | Yes |
| Data Engineering / Data Pipelines | Partial | Yes | Partial | Yes |
| Governance | No | Partial | Yes | Yes |
| Data quality | Yes | Yes | Yes | Yes |
| Metadata | Partial | Yes | Yes | Yes |
| Provenance / lineage | No | Partial | Partial | Yes |
| Versioning / reproducibility | No | Yes | Partial | Yes |
| Monitoring / observability | No | Yes | Partial | Yes |
| Explicit artifacts / documentation | Yes | Yes | Yes | Yes |
| Technology independence | Yes | Yes | Yes | Yes |

**Source:** Prepared by the authors based on Fayyad et al. (1996), Chapman et al. (2000), DAMA International (2017), Studer et al. (2021), Armbrust et al. (2021), Dehghani (2022), and Kreuzberger, Kühl and Hirschl (2023).

The comparison shows that none of the analyzed approaches was designed to fully cover the set of responsibilities considered in this study, which is consistent with their different objectives. CRISP-DM focuses primarily on the discovery and analysis process; CRISP-ML(Q) extends this scope by incorporating specific quality requirements for machine learning systems; and DAMA-DMBOK provides broader coverage of data management and governance.

The identified gap therefore lies in the systematic articulation of these responsibilities around the lifecycle of analytical assets. Among the approaches analyzed in this study, no reference structure was identified that explicitly relates technical-operational processes of Analytical Data Engineering to cross-cutting capabilities for governance, quality, metadata, provenance, versioning, reproducibility, and monitoring, while also associating them with design principles, artifacts, and evaluation mechanisms.

Given these limitations, UnespDataLens-RM is proposed as a reference model that articulates responsibilities addressed separately by different models and frameworks. Its purpose is not to replace approaches such as CRISP-DM, CRISP-ML(Q), or DAMA-DMBOK, but rather to integrate, within a single model, processes and mechanisms related to Analytical Data Engineering, governance, quality, provenance, and reproducibility.

## 3. Research Method

This research adopts Design Science Research (DSR) as the methodological approach for the design of UnespDataLens-RM. DSR is appropriate for investigations whose primary objective is to produce artifacts intended to address relevant problems by combining scientific knowledge with prescriptive design decisions (Hevner et al., 2004; Peffers et al., 2007).

According to Hevner et al. (2004), artifacts developed through DSR may take different forms, including constructs, models, methods, and instantiations. Complementarily, Gregor and Hevner (2013) emphasize that the contribution of DSR research is not limited to the artifact itself, but also includes the design knowledge associated with the decisions that guided its construction.

The adoption of DSR is consistent with the objective of this study because UnespDataLens-RM constitutes a prescriptive artifact aimed at organizing the lifecycle of Analytical Data Engineering. Its purpose is not to explain an existing phenomenon, but to specify a reference structure capable of guiding the organization of processes, responsibilities, artifacts, and control mechanisms related to the production of analytical assets.

The research program was structured according to the six activities proposed by Peffers et al. (2007): problem identification, definition of solution objectives, design and development, demonstration, evaluation, and communication. However, this article focuses on the first three activities: problem identification, definition of solution objectives, and design and development of the artifact. Demonstration and empirical evaluation constitute the planned continuation of the research and will be conducted in future work.

During the problem identification stage, concepts, models, and practices related to Data Engineering, data governance, data quality, provenance, reproducibility, DataOps, MLOps, and analytical architectures were analyzed. This analysis showed that the capabilities required to manage the lifecycle of analytical assets are distributed across approaches with different objectives and levels of abstraction, making their articulation within a common methodological structure more difficult.

During the definition of solution objectives, the general requirements intended to guide the artifact were established. These included the integration of technical-operational processes and cross-cutting capabilities, technology independence, modularity, traceability, reproducibility, and support for governance.

During the design and development stage, these requirements were translated into design principles and subsequently materialized in the UnespDataLens-RM architecture. The process resulted in the definition of technical-operational modules, cross-cutting modules, complementary dimensions, artifacts, metrics, and validation criteria. The specification was iteratively refined through consistency analysis among its components and verification of its alignment with the needs identified through the literature analysis.

## 4. Architecture of UnespDataLens-RM

UnespDataLens-RM was conceived as a reference model for organizing the lifecycle of Analytical Data Engineering through the integration of technical-operational processes and cross-cutting capabilities required for the production and evolution of analytical assets. Unlike approaches primarily oriented toward storage architecture, analytical processes, or pipeline operation, the model seeks to establish a technology-independent conceptual structure in which data, transformations, responsibilities, control mechanisms, and artifacts can be specified in an integrated manner.

The architecture results from translating the capabilities and limitations identified in the literature into design requirements. The analysis presented in Section 2 showed that analytical production processes, governance, quality, metadata, provenance, versioning, reproducibility, and monitoring are addressed by different approaches, but at different levels of abstraction and for different purposes. Based on this observation, requirements were derived to guide the organization of the artifact and, subsequently, the definition of design principles and architectural components.

Conceptually, the model is organized into three components: (i) the Technical-Operational Pipeline, which represents the activities directly related to the production of analytical assets; (ii) the Cross-Cutting Modules, which incorporate capabilities that can operate across different stages of the lifecycle; and (iii) the Complementary

Dimensions, which represent capabilities and concerns whose application depends on the context, domain, or characteristics of the analytical ecosystem. Figure 1 presents this organization and the relationships established among its components.

**Figure 1.** UnespDataLens-RM Architecture

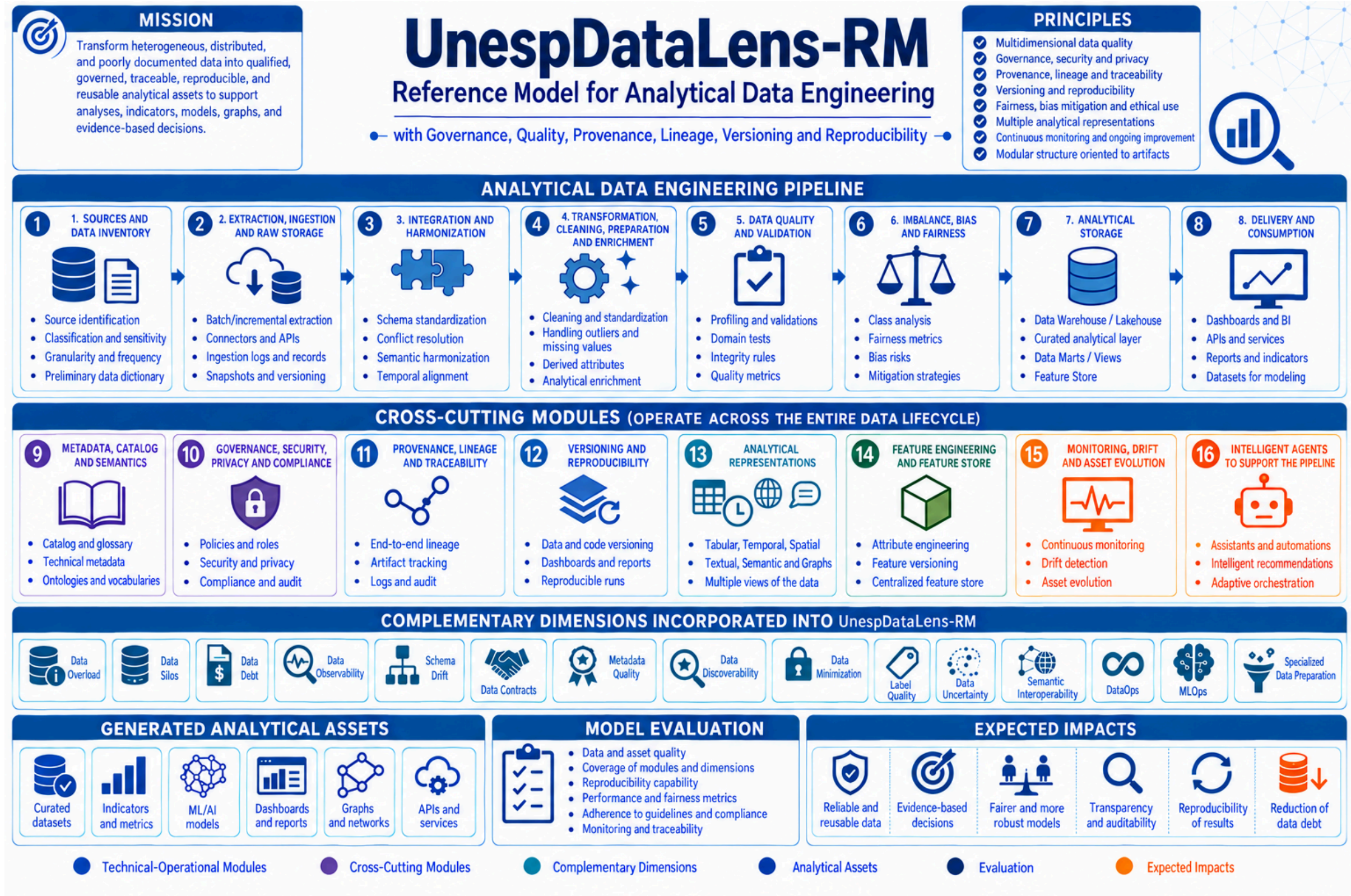

**Source:** Prepared by the authors (2026).

This separation constitutes a central design decision. Rather than representing governance, metadata, provenance, and reproducibility exclusively as sequential stages, the model positions them as capabilities that accompany the production and evolution of analytical assets throughout their lifecycle. Data quality follows a similar rationale: although the model includes a specific technical-operational module for Quality and Validation (M5), which concentrates formal activities related to assessment and conformity decisions, the Quality by Design principle establishes that quality requirements and controls should be considered continuously, from source identification to the delivery of analytical products.

## 4.1 Derivation and Design Principles

The architecture was defined through a derivation chain composed of evidence from the literature, requirements, design principles, and model components. This relationship is intended to make the rationale behind the architectural decisions explicit and to provide traceability between the investigated problem and the resulting structure.

Eight design principles were established. P1 - Data as Analytical Assets specifies that produced data should have explicitly defined purposes, documentation, responsibilities, and conditions of use. P2 - Governance by Design incorporates responsibilities, policies, classification, access, compliance, and auditing into the asset lifecycle. P3 - Quality by Design establishes that quality criteria and controls should be considered continuously rather than only after processing. P4 - Continuous Provenance and Traceability requires the recording of asset origin, transformations, dependencies, and evolution. P5 - Reproducibility requires preserving the information necessary to reconstruct processes and results. P6 - Technology Independence decouples the model from specific products, platforms, or architectures. P7 - Modularity establishes clearly delimited responsibilities and reduced coupling among components. Finally, P8 - Continuous Evolution enables the model to incorporate new capabilities and requirements without altering its structural foundations.

Table 2 summarizes the derivation of the model's main requirements. Rather than presenting the principles in isolation, the table relates the evidence identified in the conceptual background to the resulting requirements, the principles that operationalize them, and the corresponding architectural components.

**Table 2.** Derivation of UnespDataLens-RM Requirements and Components

| Identified evidence | Derived requirement | Principle(s) | Related components |
|---|---|---|---|
| Fragmentation of activities across the analytical production lifecycle | Organize the lifecycle of analytical assets into explicitly defined processes | P1, P7 | M1-M8 |
| Heterogeneity of data sources, structures, and technological environments | Enable data acquisition, integration, and processing independently of specific technologies | P6, P7 | M1-M4 |
| Multidimensional data quality dependent on the context of use | Incorporate quality criteria and controls throughout the lifecycle | P3 | M1-M8, with consolidation in M5 |
| Governance frequently separated from pipeline execution | Incorporate policies, responsibilities, access controls, and compliance throughout the lifecycle | P2 | M10 |
| Dependence on documentation for understanding and reuse | Maintain a structured description of analytical assets and their context | P1, P7 | M9 |
| Loss of information about data origin, transformations, and dependencies | Maintain end-to-end traceability of analytical assets | P4 | M11 |
| Changes in data, code, and configurations may compromise the reconstruction of results | Preserve versions and the information required to reproduce processes and results | P5 | M12, in articulation with M9 and M11 |
| Different analytical products require specific representational structures | Support multiple analytical representations without altering the reference process | P6, P7 | M13 |
| Certain analytical contexts require reusable derived attributes | Support the management of features when required by the analytical product | P1, P7 | M14 |
| Pipelines and data distributions change over time | Continuously monitor the behavior, stability, and evolution of analytical assets | P8 | M15 |
| Analytical ecosystems evolve as new requirements and capabilities emerge | Allow extensions without compromising the conceptual structure of the model | P6, P7, P8 | Complementary Dimensions and extensions |
| Increasing use of intelligent mechanisms for accessing, documenting, and retrieving contextual information | Enable assisted consultation, documentation, and recommendation mechanisms without making them mandatory components of the model core | P7, P8 | M16 |

**Source:** Prepared by the authors (2026).

The derivation shows that the model components were not defined merely by aggregating existing concepts, but by transforming needs identified in the literature into architectural requirements. These requirements guide the design principles, which in turn determine how responsibilities are distributed across technical-operational processes, cross-cutting capabilities, and model extensions. This traceability also establishes a basis for future evaluations, in which the consistency and coverage of the artifact can be assessed against the requirements that guided its development.

## 4.2 Structure and Organization of the Modules

The UnespDataLens-RM architecture comprises sixteen modules, divided into eight technical-operational modules and eight cross-cutting modules. The former represent the main production flow of analytical assets, whereas the latter provide capabilities that can operate at different points throughout this flow. This organization seeks to separate operational responsibilities from management and control responsibilities while preserving their interaction.

The technical-operational modules structure the progression of data from source identification to delivery for consumption. Although they are presented in a logical sequence, their execution does not need to follow a strictly linear path. Depending on the instantiation, activities may be iterative, performed in parallel, or return to previous modules as a result of validation outcomes, changes in requirements, or problems identified during processing. Table 3 presents their objectives, main inputs, and main outputs.

**Table 3.** Technical-Operational Modules of UnespDataLens-RM

| Module | Objective | Main inputs | Main outputs |
|---|---|---|---|
| **M1 - Sources and Inventory** | Identify, catalog, and characterize internal and external data sources. | Systems, databases, APIs, files, sensors, and documents. | Source inventory, classification, and initial metadata. |
| **M2 - Extraction and Ingestion** | Perform controlled data acquisition and ingestion while preserving the original state of the data. | Inventoried sources and collection configurations. | Raw data, ingestion records, logs, and snapshots. |
| **M3 - Integration and Harmonization** | Integrate data from different sources and address structural and semantic incompatibilities. | Data obtained from the sources. | Integrated data, matching results, integration rules, and mappings. |
| **M4 - Transformation and Preparation** | Clean, standardize, enrich, and prepare data for analytical use. | Integrated data. | Processed data and transformation records. |
| **M5 - Quality and Validation** | Evaluate analytical assets according to defined metrics, rules, and validation criteria. | Prepared data and validation criteria. | Quality indicators, non-conformities, and approval or rejection decisions. |
| **M6 - Bias and Fairness** | Assess representativeness, imbalance, and potential biases in the data. | Validated data and information about relevant groups. | Distribution indicators, bias analyses, and documented limitations. |
| **M7 - Analytical Storage** | Maintain analytical assets in structures suitable for consumption requirements. | Approved assets. | Persistent analytical structures and associated policies. |
| **M8 - Delivery and Consumption** | Make analytical assets available to users, applications, and analytical processes. | Stored assets. | Data products, APIs, dashboards, curated datasets, and documentation. |

**Source:** Prepared by the authors (2026).

The explicit definition of inputs and outputs establishes conceptual interfaces between the modules and allows their implementation to be adapted to different technologies. For example, the model does not prescribe whether M7 should use a data warehouse, data lake, or lakehouse; rather, it specifies the responsibilities that an instantiation must fulfill to properly persist and make the produced assets available.

The cross-cutting modules, in turn, represent capabilities whose operation is not restricted to a specific position in the flow. They may produce or consume information from different technical-operational modules while accompanying the evolution of analytical assets throughout their lifecycle. Table 4 summarizes their organization.

**Table 4.** Cross-Cutting Modules of UnespDataLens-RM

| Module | Objective | Scope of operation | Main artifacts |
|---|---|---|---|
| **M9 - Metadata and Catalog** | Organize descriptive, technical, and operational information about analytical assets. | Continuous documentation of assets and their relationships. | Catalog, glossary, data dictionary, and asset information sheets. |
| **M10 - Governance and Compliance** | Define responsibilities, policies, and controls. | Access control, classification, compliance, and accountability. | Responsibility matrix and access and classification policies. |
| **M11 - Provenance and Lineage** | Record the origin, transformations, and dependencies of analytical assets. | Traceability throughout the pipeline. | Lineage graph, dependency records, and impact analyses. |
| **M12 - Versioning and Reproducibility** | Preserve versions and the conditions required to reconstruct analytical processes. | Control the evolution of data, code, rules, and computational environments. | Manifests, reproducibility packages, changelogs, and version records. |
| **M13 - Analytical Representations** | Structure data according to consumption and analysis requirements. | Design and evolution of analytical products. | Tabular, temporal, and spatial representations, graphs, documents, and semantic structures. |
| **M14 - Feature Engineering** | Manage derived attributes when required by analytical applications. | Preparation and evolution of products that use features. | Feature catalog, feature store, and documentation. |
| **M15 - Monitoring and Drift** | Monitor the behavior, stability, and evolution of assets and pipelines. | Continuous operation and evolution. | Indicators, drift alerts, reports, and monitoring logs. |
| **M16 - Intelligent Agents** | Support consultation, documentation, recommendations, and maintenance when applicable. | Assisted support for lifecycle activities. | Agents, recommendations, interaction histories, and human-validation records. |

**Source:** Prepared by the authors (2026).

The cross-cutting modules should not be interpreted as requiring uniform implementation. The intensity and form of their application depend on the characteristics of each instantiation. This is particularly relevant for M13, M14, and M16: different analytical representations may be required depending on the analytical product; Feature Engineering is primarily applicable when derived attributes are needed for analyses or models; and intelligent agents constitute a supporting capability that may be incorporated when appropriate, without being a necessary condition for conformity with the model.

The modular specification of UnespDataLens-RM also supports alternative forms of component instantiation. The explicit definition of responsibilities, inputs, outputs, artifacts, rules, metrics, and validation criteria establishes conceptual interfaces that allow specific functions to be implemented either by conventional software components or, when appropriate, by specialized artificial intelligence agents. From this perspective, an agent may assume responsibilities associated with a module or with a specific set of activities, interacting with other components through the artifacts and interfaces defined by the model. This possibility is not a requirement for adopting UnespDataLens-RM, but rather a consequence of its modularity and technology independence, enabling future instantiations based on multi-agent architectures.

### 4.3 Complementary Dimensions and Extensibility

In addition to the modules that constitute its main structure, UnespDataLens-RM includes complementary dimensions intended to represent practices, concerns, and capabilities that cut across different components or whose application depends on the characteristics of the environment. These dimensions include DataOps, MLOps, Data Contracts, observability, Data Discoverability, semantic interoperability, and Schema Drift management.

These dimensions do not constitute additional pipeline stages or necessarily independent modules. Instead, they function as mechanisms for specialization and extension of the model. DataOps and MLOps, for example, may influence automation, versioning, deployment, and monitoring; Data Contracts may establish constraints between data producers and consumers; and semantic interoperability may extend capabilities related to integration, metadata, and analytical representations.

This separation also supports the evolution of UnespDataLens-RM. New practices or capabilities may initially be incorporated as complementary dimensions before sufficient theoretical and empirical evidence justifies their formalization as structural components. In this way, the model seeks to preserve stability in its conceptual core while remaining adaptable to the evolution of technologies and practices in Analytical Data Engineering.

Taken together, the architecture establishes an explicit relationship among evidence from the literature, design requirements, principles, technical-operational processes, and cross-cutting capabilities. This structure constitutes the main mechanism through which UnespDataLens-RM seeks to integrate responsibilities currently distributed across different approaches while preserving modularity, technology independence, and the possibility of specialization.

## 5. Discussion

UnespDataLens-RM was conceived as a reference model for organizing the lifecycle of Analytical Data Engineering through the integration of technical-operational processes and cross-cutting capabilities related to governance, quality, provenance, versioning, reproducibility, analytical representations, monitoring, and intelligent support. The main contribution of the proposal does not lie in the isolated introduction of these concepts, since many of them are already well established in the literature, but rather in their systematic articulation within a common methodological structure. This characteristic positions the model as an integrating mechanism across different areas that, although complementary, are often addressed independently in process models, data architectures, and management practices.

The first contribution of UnespDataLens-RM lies in the formalization of a lifecycle for Analytical Data Engineering. Classical models such as KDD and CRISP-DM were fundamental to systematizing knowledge discovery and data mining processes (Fayyad et al., 1996; Chapman et al., 2000), while more recent approaches such as CRISP-ML(Q), DataOps, and MLOps have broadened the focus to include quality, operationalization, automation, and monitoring of analytical solutions. UnespDataLens-RM shifts the focus toward the methodological infrastructure required for the production and maintenance of analytical assets themselves, covering activities from the identification and characterization of data sources to the delivery of assets for consumption.

A second contribution consists of integrating cross-cutting capabilities into the technical-operational pipeline. Governance, metadata, provenance, versioning, and monitoring are not treated as subsequent stages or independent activities, but as capabilities that continuously accompany analytical assets throughout their lifecycle. This organization seeks to bring established principles of data governance, provenance, reproducibility, and modern pipeline management into a common conceptual structure. Within the complete model, these capabilities are further articulated with mechanisms for quality, feature engineering, analytical representations, observability, and other complementary dimensions.

The third contribution concerns the formalization of the model components through technical-methodological artifacts. Rather than defining only abstract stages, UnespDataLens-RM associates its modules with intermediate products capable of documenting decisions, rules, metrics, transformations, and validation conditions.

A fourth contribution derives from the adoption of Governance by Design, Quality by Design, traceability, and reproducibility as guiding principles for pipeline development. From this perspective, the quality of an analytical product does not depend exclusively on the outcome of its final stage, but also on the ability to understand how its data were acquired, transformed, evaluated, and delivered. This perspective is intended to support auditing, impact analysis, result reproduction, and the understanding of dependencies among analytical assets.

Finally, technology independence and modularity increase the model's potential for specialization. UnespDataLens-RM does not prescribe specific databases, ETL/ELT tools, orchestration platforms, or analytical environments. Its modules represent conceptual responsibilities that may be materialized through different technologies. Consequently, an organization may instantiate only those components that are compatible with its level of maturity and progressively expand the adoption of the model without requiring the simultaneous implementation of all proposed capabilities.

The contributions and limitations discussed above indicate that, at the current stage of the research, UnespDataLens-RM should be understood as a technical-methodological proposition undergoing validation, rather than as a model whose superiority over existing approaches has already been empirically demonstrated. Its central contribution lies in the systematization and integration of responsibilities that are typically distributed across different Data Engineering models, frameworks, and practices.

## 6. Conclusions

This study presents UnespDataLens-RM, a reference model for Analytical Data Engineering designed to integrate technical-operational processes with capabilities related to governance, quality, provenance, lineage, versioning, and reproducibility. The proposal is based on the understanding that the reliability of analytical assets depends not only on data acquisition, transformation, and storage operations, but also on the ability to document their origin, control their evolution, assess their quality, and reconstruct the processes responsible for their production.

Developed according to Design Science Research, the model was structured into technical-operational modules, cross-cutting modules, complementary dimensions, design principles, and artifacts associated with the analytical lifecycle. Its main contribution lies in articulating, within a single conceptual structure, responsibilities that are typically addressed separately across process models, governance frameworks, Data Engineering practices, and operational approaches. UnespDataLens-RM is not intended to replace these proposals, but rather to provide a reference structure through which they can be related and specialized according to the application context.

Technology independence and the possibility of specialization across different domains allow the model components to be materialized in different infrastructures and at different levels of organizational maturity. In addition, the formalization of the modules through artifacts, metrics, and validation criteria seeks to bring the conceptual specification closer to future operational instantiations.

The results presented here should, however, be interpreted in light of the current stage of the research. UnespDataLens-RM is at an advanced level of conceptual specification, whereas its demonstration and empirical evaluation are still to be conducted. Therefore, the present study does not allow us to conclude that its use necessarily produces measurable improvements in quality, governance, traceability, or reproducibility, nor does it establish superiority over other forms of pipeline organization.

Future work will focus on four main directions: (i) developing a reference instantiation and experimentally evaluating the model; (ii) investigating incremental adoption strategies and defining conformance or maturity levels; (iii) applying the model across different domains to assess its generality and the need for specialization; and (iv) further investigating interoperability with existing standards and approaches, including W3C PROV, FAIR, DataOps, and MLOps, as well as instantiation architectures in which responsibilities defined by the modules may be performed by specialized and coordinated artificial intelligence agents.

In summary, UnespDataLens-RM establishes a conceptual foundation for treating Analytical Data Engineering as a governed lifecycle of analytical assets. The consolidation of its contribution will depend on subsequent stages of instantiation, experimentation, and evaluation in real-world scenarios, which should determine the extent to which the proposed integration effectively contributes to the construction of more reliable, traceable, auditable, and reproducible analytical ecosystems.

CRediT Authorship Contribution Statement

Ronaldo Celso Messias Correia, Douglas Francisquini Toledo, and Camila Tolin Santos da Silva jointly contributed to the conceptualization of the research, methodology definition, investigation, formal analysis, conception and development of UnespDataLens-RM, preparation of the model representations, drafting of the original manuscript, and review and editing of its final version. All authors participated in the critical analysis of the content, approved the final version of the manuscript, and take responsibility for the work presented.

Funding

This study was financed in part by the Coordenação de Aperfeiçoamento de Pessoal de Nível Superior – Brasil (CAPES) – Finance Code 001.

## Declaration of Competing Interests

The authors declare that they have no competing financial interests or personal relationships that could have influenced the work reported in this article.

## Declaration of Generative AI and AI-Assisted Technologies in the Manuscript Preparation Process

During the preparation of this manuscript, the authors used ChatGPT, developed by OpenAI, as a tool to support text organization, linguistic revision, writing improvement, adjustment of the article structure, and image production. The tool was used under the authors' supervision and did not replace critical analysis, methodological decision-making, or the scientific responsibility of the researchers. All content produced or revised with the assistance of the tool was subsequently analyzed, verified, and edited by the authors, who take full responsibility for the accuracy, originality, and content of the final version of the manuscript.

## Data Availability

This study presents the design and specification of a reference model and, at this stage of the research, did not involve the generation or empirical analysis of datasets. Therefore, no datasets are associated with the results presented in this article.